\documentstyle[preprint,aps]{revtex}
\begin{document}
\newif\ifplot
\plottrue
\newcommand{\RR}[1]{[#1]}
\newcommand{\intsum}{\sum \kern -15pt \int}
\newfont{\Yfont}{cmti10 scaled 2074}
\newcommand{\Y}{\hbox{{\Yfont y}\phantom.}}
\def\O{{\cal O}}
\newcommand{\bra}[1]{\left< #1 \right| }
\newcommand{\braa}[1]{\left. \left< #1 \right| \right| }
\def\Bra#1#2{{\mbox{\vphantom{$\left< #2 \right|$}}}_{#1}
\kern -2.5pt \left< #2 \right| }
\def\Braa#1#2{{\mbox{\vphantom{$\left< #2 \right|$}}}_{#1}
\kern -2.5pt \left. \left< #2 \right| \right| }
\newcommand{\ket}[1]{\left| #1 \right> }
\newcommand{\kett}[1]{\left| \left| #1 \right> \right.}
\newcommand{\scal}[2]{\left< #1 \left| \mbox{\vphantom{$\left< #1 #2 \right|$}}
\right. #2 \right> }
\def\Scal#1#2#3{{\mbox{\vphantom{$\left<#2#3\right|$}}}_{#1}
{\left< #2 \left| \mbox{\vphantom{$\left<#2#3\right|$}}
\right. #3 \right> }}
\draft
\title{
Resonances in the three-neutron system
}
\author{H. Wita\l a$^*$, W. Gl\"ockle
}
\address{
Institut f\"ur theoretische Physik II, Ruhr-Universit\"at Bochum,
D-44780 Bochum, Germany
}
\address{$^{*}$ Institute of Physics, Jagellonian University, 
PL- 30059 Cracow, Poland}

\date{\today}
\maketitle
\widetext
\begin{abstract}
A study of 3-body resonances has been performed in the framework of 
configuration space Faddeev equations. 
The importance of keeping a sufficient number of terms in the 
asymptotic expansion of the resonance wave function is pointed out. 
We investigated three neutrons interacting in selected force 
components  taken from realistic nn forces. Three-neutron resonance pole 
trajectories connected to artificially enhanced nn forces could be found. 
The final pole positions corresponding to the actual force strengths could not 
be found due to the onset of numerical instabilities. The numerically reliable 
results, however, achieved makes it likely, that three-neutron 
resonance energies will have large imaginary parts and therefore physically 
not interesting. A straithforward application of complex scaling 
to 3-nucleon systems with realistic forces could not be controlled 
numerically. 
\end{abstract}

\pacs{ PACS numbers: 21.30.+y, 21.45.+v, 24.10.-i, 25.10.+s}
\pagebreak
\narrowtext

\section{Introduction}

Three-nucleon (3N) resonances have not yet been firmly established. The
situation up to 1987 has been collected in ~\cite{til87}. 
 A recent investigation ~\cite{yul97} of the process 
 $^3He(\pi^-,\pi^+)3n$ 
found no evidence of a resonance state of three neutrons. Earlier
investigations ~\cite{sper70}, ~\cite{wil69} 
 pointed to resonances in the
 three neutron (3n)  and three proton (3p) systems. 
The double charge exchange process on $^3He$ was also 
investigated in ~\cite{set86} criticizing 
previous work but pointing again to a 3n  resonance, now
around $(20 - i~20)$~MeV. For the  nnp system an excited state around $7$~MeV 
with a width of $\Gamma = 0.6 \pm 0.3$~MeV has been suggested in 
~\cite{ale94}. A variational study based on complex scaling and
simplified nucleon-nucleon (NN) forces was carried through in ~\cite{cso96} 
 with the prediction of a 3n resonance at
$E=(14 - i~13)$~MeV for a $J^{\pi}=3/2^+$ state. Earlier theoretical
studies ~\cite{4}, ~\cite{glo78}  
 might be useful in various aspects, but are not 
conclusive with respect to the actual position of 3n resonances. A
virtual state of $^3H$ with the quantum numbers of the ground state has 
been investigated in ~\cite{del84} 
 by analytically continuing the momentum space Faddeev equation and
even estimating three-nucleon force (3NF) effects. 
In the framework of the hyperspherical 
harmonic method and using complex scaling with model potentials a
subthreshold 3n resonance for $J^{\pi}=1/2^-$ has been located in 
~\cite{sof97}. Recently mathematical foundations have 
been laid on the analytical continuation of the three-body Faddeev 
equations into unphysical energy sheets ~\cite{mot97} 
 and an application thereof to the $^4He$ trimer 
appeared in ~\cite{motkol97}. 

It is a theoretical challenge to determine the locations of the lowest
lying 3N resonances on unphysical sheets based on the most
modern nuclear forces. Those forces are very successful in describing
3N scattering observables~\cite{5a} which leads to the
expectation that the corresponding 3N Hamiltonian will also locate 3N
resonances with a high degree of reliability. Like for the 3N bound
state energy, where some theoretical underbinding results and 
3NF's are used to
bridge that gap to the data point, also 3N resonance positions might
finally be fine-tuned in this manner. These future rigorous results would 
also be an important guidance for experiments and would clarify the
hitherto controversial situation.

For the 3N system the  bound state has been studied 
extensively in the last decades resulting in numerically precise and 
converged solutions of the 3N Faddeev 
equations~\cite{4a},~\cite{chen86},~\cite{carl98},~\cite{viv98},~\cite{wu93}. 
 Energy eigenvalues 
and wave functions have been generated with most modern nuclear 
Hamiltonians in momentum and coordinate space. Also numerically exact 
solutions of 3N Faddeev equations for continuum states with the same 
Hamiltonians are now available in momentum space for a wide range 
of energies~\cite{5a}. The first model 
solutions also in coordinate space appeared~\cite{6} and  
variational approaches matured and provide equally accurate 
solutions~\cite{pisa98}. 

3N resonances, however,  have 
not yet been investigated  in a comparable 
rigorous manner and based on realistic NN forces 
to the best of our 
knowledge. In this article we would like to take first steps into such 
a direction. 

In the two-nucleon (2N) system the exact 
form of the wave function in configuration space 
is known outside the range of the potential. In the 3N system only asymptotic 
expansions are known~\cite{5},~\cite{glo74} 
and methods how to apply them with sufficient accuracy, for instance in 
Faddeev equations, have still to be elaborated in 
that framework. There are first very promising results for 3N scattering 
above the 3N breakup threshold~\cite{6} though up to now only for simple 
S-wave model forces. To the best of our knowledge 3N resonance wave functions, 
which asymptotically oscillate and grow exponentially, have not yet 
been determined as solutions of Faddeev equations and realistic 
forces. First attempts in this 
direction have been done in ~\cite{10}, however with very simple forces. 

In this study we investigate 3-body resonances as solutions of configuration 
space Faddeev equations. The largest obstacle to find the resonance positions 
with sufficient precision is the approximate form of the resonance boundary 
condition. Ab initio it is not known at which coordinate values this 
form is valid with sufficient accuracy. For resonances this problem 
can be overcome in principle using the complex 
scaling method (CSM)~\cite{15},~\cite{18}. 
Thereby all relative coordinates are ``rotated'' into the complex 
plane leading to bound-state like, exponentially decreasing resonance 
wave function. This ``rotation'' of the coordinates into the complex 
plane works well  for problems in atomic physics with potentials 
of inverse power laws. It will be shown in this article that  in 
nuclear physics  
a straighforward application of 
 CSM in the configuration space Faddeev form leads to serious 
numerical problems 
when the usual realistic NN 
interactions of Yukawa type  and hard core 
behavior are used. Whether feasible  modifications thereof can be found 
remains to be seen. Right now our experience is such, 
 that the resonance energies can be gained 
with sufficient precision only if  
one does not leave the real r-axis and if one guaranties 
 the exact boundary condition as close as possible by using 
a sufficient number of terms in 
its 
asymptotic form. 

In section II we explain in detail the problems  encountered when 
one applies the CSM to the nuclear resonances using the 2N 
system as a play ground. The exact knowledge of the boundary conditions 
for this system allows one to check assumptions which can  later be applied to
3N resonances. The 3N equations and the method for their solution are 
presented 
in section III. There we discuss results for 3 bosons interacting with 
simple, S-wave forces and more importantly for 3n's interacting 
via realistic neutron-neutron (nn) interactions. 
A summary is given in section IV. Information on the technical performance 
can be found in the Appendices A and B.

\section{Resonances in the 2N system}

Searching  resonances in a system of two-nucleons 
 is a well defined problem and 
numerically well under control~\cite{gl83}. 
In spite of that we discuss it here and 
present some results due to its pedagogical simplicity, which 
enables us to present
 and test approaches which might be useful for finding resonances for 
three or more nucleons. 
 In addition we can  introduce our notation. 

The 2N resonances are associated with the poles of the S-matrix which 
are embedded in the fourth-quadrant of the complex k-plane (Re(k)$>0$, 
Im(k)$<0$). They are solutions of the time-independent Schr\"odinger 
equation without  incoming wave and the outgoing wave increasing 
exponentially  at infinity. We introduce the coordinate space 
partial wave basis $\vert r,\alpha_2> \equiv \vert r,(ls)jt >$ 
where  
\begin{eqnarray}
< \overrightarrow {r}' \vert rlm > = {{\delta(r'-r)}\over{rr'}} Y_{lm}(\hat r')
\label{eq1}
\end{eqnarray} 
and where the orbital angular momentum l and total spin s are coupled 
to the total angular momentum j. The isospin of the two nucleons is 
denoted by t. 
Then the resonance wave function  $\psi(\overrightarrow {r})$ to a 
given j and t has the form
\begin{eqnarray}
 \psi(\overrightarrow {r}) = \sum_l{{{\phi_{\alpha_2}(r)}\over{r}} }
 < \hat r \vert \alpha_2 > .
\label{eq2}
\end{eqnarray} 
It is written in terms of the reduced wave functions $\phi_{\alpha_2}(r)$ 
which fulfill the radial Schr\"odinger equation 
\begin{eqnarray}
\sum_{\alpha_2'} \lbrack \delta_{\alpha_2 \alpha_2'} 
( {{d^2}\over{dr^2}} - {{l_{\alpha_2}( l_{\alpha_2}+1 )}\over{r^2}} 
 ) - {m\over{{\hbar}^2}}V_{\alpha_2 \alpha_2'}(r) \rbrack \phi_{\alpha_2'}(r) 
= 
-k^2\phi_{\alpha_2}(r) .
\label{eq3}
\end{eqnarray}

The resonance conditions are such that $\psi$ has to be regular at the origin 
and purely outgoing. For the reduced amplitudes  this amounts to 
\begin{eqnarray}
\phi_{\alpha_2}(0)=0
\label{eq4}
\end{eqnarray}
and 
\begin{eqnarray}
\phi_{\alpha_2}(r) \propto  r h^{(1)}_{l_{\alpha_2}}(kr)
\label{eq5}
\end{eqnarray}
outside the range of the interaction. Here $h^{(1)}_{l}(kr)$ is the 
spherical Hankel 
function. 
As we shall demonstrate below it is mandatory to either use the exact form 
of the boundary condition (\ref{eq5}) or to take into account at least 
the leading orders 
 to find the resonances in the 2N system with sufficient accuracy. 
In Eq.(\ref{eq3}) m is the nucleon mass and $k^2 \equiv {m\over{{\hbar}^2}}E$. 
The complex energy E is embedded in the lower half plane of the second 
energy sheet. The boundary conditions (\ref{eq4}) and (\ref{eq5}) 
apply also to bound states with real, negative eigenvalues E.
The matrix elements of the NN interaction are defined in terms of 
$V_{\alpha_2 \alpha_2'}(r)$ by 
\begin{eqnarray}
< r \alpha_2 \vert V \vert r' \alpha_2' > = 
{{\delta(r-r')}\over{rr'}} V_{\alpha_2 \alpha_2'}(r)
\label{eq6}
\end{eqnarray}

A method of solving Eq.~(\ref{eq3}) is  displayed 
in~\cite{8}. One chooses suitable grid points in $r$  
and interpolates $\phi_{\alpha_2}(r)$ in terms of cubic splines. 
The boundary conditions (\ref{eq4}) and (\ref{eq5}) 
can be conveniently incorporated 
either by the absence of certain expansion coefficients in the spline 
expansion or by suitable relations among them. Then choosing 
appropriate collocation points in $r$  one ends up with a 
homogeneous algebraic set of equations. 
They can be handled by standard methods~\cite{8}.

In systems with more than 2 nucleons the boundary 
conditions are known only via an asymptotic expansion and one has therefore 
 to work 
with approximate boundary conditions. 
The knowledge of the exact resonance solutions  in the 2N system allow 
now to study 
the question pivotal for 3N systems: how precisely does one need to know 
the boundary conditions in order to get the resonance 
energies with sufficient accuracy? 
 To study this problem one 
can use the series representation for the spherical Hankel functions
\begin{eqnarray}
h^{(1)}_l(x) = \sum_{m=0}^l {{(l+m)!}\over{m!(l-m)!}}{1\over{2^m}}
{{exp(i\lbrace x - {{l-m+1}\over{2}} \pi \rbrace)}\over{x^{m+1}}} 
\label{eq7}
\end{eqnarray}
and write the resonance wave function outside the interaction region 
in a form which  
resembles the asymptotic 
boundary condition  in the 3N system (see Section III, Eq.(\ref{eq21}))
\begin{eqnarray}
\phi(r) =  exp(ikr) ~ \sum_{m=0}^l {{a_m}\over{(kr)^m}}
\label{eq8}
\end{eqnarray}
with $a_m$ being unknown constants independent from k and r. 
Here, contrary to the 3N system, the number of terms in the series is 
 restricted by 
the orbital angular momenta contributing to the given total angular momentum 
j of a resonance. 

In Table~\ref{table1} 
we present 2N resonance positions in different NN states where the 
 exact boundary condition have been imposed at different 
matching radii outside the interaction region. The NN interaction 
 Reid93~\cite{r93} has been used and has been multiplied with a factor 
$\lambda_V$ (shown in column 2 of Table~\ref{table1}). 
 Using the  exact boundary condition the 
resonance positions do not depend of course 
on the value of the matching radius. 
However, when the most crudely approximated boundary condition 
is applied, keeping in Eq.~(\ref{eq8}) only the term with $m=0$, 
the approximate resonance positions vary with 
the matching 
radius and up to $20$~fm the results are highly insufficient. 
As seen from Table~\ref{table1} it is 
sufficient  to get correct energy eigenvalues 
if one keeps only the two leading terms in the boundary condition 
(\ref{eq8}) ( for l=1 this is the exact form). 

For resonances the problem of a possibly approximate nature of the boundary 
condition can be overcome in principle using 
the complex scaling method (CSM)~\cite{15},~\cite{18}. 
This will now be illustrated for 
2 nucleons because of its simplicity.  
In the 2N system one replaces the relative coordinate r  by 
\begin{eqnarray}
r \rightarrow  r' \equiv r~e^{i\alpha} .
\label{eq9}
\end{eqnarray}  
For real positive angles $\alpha$ the asymptotic form 
$e^{ik\rho}$ with $Im(k) < 0$ can be changed to an 
exponentially decreasing function. Therefore a resonance wave function  
analytically continued to those complex coordinates can be treated 
like a bound state with its well known boundary condition. 
Then it is probably enough to restrict  the boundary 
condition (\ref{eq8}) to the first term $m=0$.  
This ``rotation'' of the coordinate into the 
complex plane is sufficient for problems in atomic physics with 
Coulomb potentials. In nuclear physics the potentials 
decrease exponentially and the transformation (\ref{eq9}) turns then 
$e^{-\mu~r}$ into $e^{-\mu(cos(\alpha) + isin(\alpha))r}$. Therefore 
for $\alpha > 0$ the range of the potential will increase, which is 
counterproductive. This can be avoided by generalizing (\ref{eq9}) to
complex $\alpha$'s. Then 
\begin{eqnarray}
e^{-\mu~r} \rightarrow exp \lbrace -\mu e^{-Im(\alpha)} \lbrack 
cos( Re(\alpha)) + 
i sin (Re(\alpha)) \rbrack r \rbrace 
\label{eq10}
\end{eqnarray}
which for $Im(\alpha) < 0$ increases the mass $\mu cos(Re(\alpha))$ 
in the exponent to 
$\mu e^{-Im(\alpha)}cos(Re(\alpha))$ and thus choosing $Im(\alpha)$ 
suitably decreases the range of the 
interaction again. 
At the same time also the wave function decreases faster due to 
the 
factor $e^{-Im(\alpha)}$ in the exponent. Let us call that transformation 
extended CSM (ECSM). 

We firstly illustrate  the power of the 
ECSM in the 2N system using the Reid~93 
NN potential in the state $^3P_0$. As we have seen 
in Table~\ref{table1}, an enhancement factor
 $\lambda_V=5.5$ leads  to 
a resonance state with a complex energy eigenvalue $E=(1.42 - i~0.75)~$MeV.
If one uses in Eq.(\ref{eq8}) the first  term with $m=0$ only 
the energy eigenvalue is badly presented 
(see Tables~\ref{table1}  and  \ref{table2}). 
The result is  improved and we can gain the correct result with 
the required precision using CSM with $Im(\alpha) = 0$ 
(see Table~\ref{table2}). 
Note that in Table~\ref{table2} 
we stick to the asymptotic form ($m=0$) in view of the fact 
that for 3 bodies we also have only approximate forms. 
 If we now allow for negative $Im(\alpha)$'s then as can be seen 
in Table~\ref{table2} we gain 
the correct eigenvalue even choosing a very small matching radius.

We also see from Table~\ref{table2} that CSM for increasing $Re(\alpha)$ 
gets numerically unstable at a certain value of $Re(\alpha)$. 
In this example it happens around $40^o$. 
   This behavior 
is intricately connected to the present day parametrizations 
of  NN interactions and puts into question the applicability of 
 CSM for finding broad resonances in  nuclear systems. 
The parametrizations are typically of Yukawa type (one boson 
exchanges) like in the Reid93~\cite{r93} or 
Reid soft core potential~\cite{rsc}.
The analytical continuation to complex r-values 
leads  to an oscillating behavior of the potential, which drastically 
increase with increasing $Re(\alpha)$, 
with unstable numerics as a consequence. 
We examplify this behavior in Fig.~\ref{fig1} for 
the Reid93 potential. We checked several of the existing NN potentials 
and found always this problem ( potentials according to Gogny~\cite{gogny}, 
R.de Tourreil and D.W.L. Sprung ~\cite{dts}, 
 H. Eikemeier and H.H. Hackenbroich~\cite{hack}, and 
  R. de Tourreil~\cite{dt}). 
In case of the AV14~\cite{av14} and AV18~\cite{av18} potentials 
the situation is even worse due to a singularity  for complex r-values 
in their 
parametrization of the hard core part. Also Gauss-parametrizations are not 
an exception. In this case the form 
$exp(-ar^2)$ limits  the complex scaling angle $Re(\alpha)$ to 
$Re(\alpha)<45^o$.

One possibility to generalize the CSM in order to avoid the above 
problems is 
the so called ``smooth-exterior'' complex scaling (SECS)~\cite{18}. 
The idea is to perform the complex rotation only outside the region 
of interaction. In this way the 
analytical continuation of the potential into the complex 
coordinate plane is avoided. In practice this can 
be achieved for instance by performing the following 
transformation (note the printing errors in~\cite{18} leading to 
false expressions (2.1.30) and (2.1.32)) 
\begin{eqnarray}
r \rightarrow  r' &=& F(r)\to r~e^{i\alpha}~~~ for~~~ r \to \infty
\label{eq11}
\end{eqnarray}
with 
\begin{eqnarray}
F(r) = r + \lbrack exp(i\alpha)-1 \rbrack \lbrace r + {{1}\over{4\lambda}} 
ln {{ \lbrack 1 + exp(2\lambda(r-r_0)) \rbrack  
\lbrack 1 + exp(-2\lambda(r-r_0)) \rbrack }\over{ 
\lbrack 1 + exp(2\lambda(r+r_0)) \rbrack  
\lbrack 1 + exp(-2\lambda(r+r_0)) \rbrack}} \rbrace .
\label{eq12}
\end{eqnarray}
This transformation  leads to the smooth-exterior-scaled Hamiltonian 
\begin{eqnarray}
H &=& -{{\hbar^2}\over{m}} {{1}\over{f^2(r)}} {{d^2}\over{dr^2}} + V( F(r)) + 
       {{\hbar^2}\over{m}} {{f'(r)}\over{f^3(r)}} 
       {{d}\over{dr}} + {{\hbar^2}\over{m}} {{l(l+1)}\over{F^2(r)}}
\label{eq13}
\end{eqnarray}
with 
\begin{eqnarray}
f(r) &=& {{dF}\over{dr}} = 1 + \lbrack exp(i\alpha) -1 \rbrack 
g(r) \nonumber \\
g(r) &=& 1 + 0.5 \lbrack tanh(\lambda(r-r_0)) -  
tanh(\lambda(r+r_0)) \rbrack 
\label{eq14}
\end{eqnarray}
and 
\begin{eqnarray}
{{df}\over{dr}} &=& 0.5\lambda \lbrack exp(i\alpha) -1 \rbrack  \lbrack 
cosh^{-2}(\lambda(r-r_0)) - cosh^{-2}(\lambda(r+r_0)) \rbrack .
\label{eq15}
\end{eqnarray}
The case $\lambda=0$ corresponds to usual CSM  $r' = r~e^{i\alpha}$~\cite{18}. 
Taking $r_0$ outside the region of the 
NN interaction and choosing a proper value 
for $\lambda$ a smooth-exterior-scaling path results, which avoids the 
oscillating 
changes of the NN potential.  We present in  Table~\ref{table3} 
the effectiveness of this approach by  
 aplying it to the $^1D_2$ resonance state for the  Reid93 NN 
potential with $\lambda_V=5.0$. It is enough to keep only the first 
term in the boundary condition (\ref{eq8}) to reproduce the 
correct resonance position when a  sufficiently large 
value of the scaling angle $\alpha$ is chosen. 
Now, however, contrary to CSM, no numerical 
problems arise when large values of $\alpha$ are used. In spite of its 
success in the 2N system we think that the 
SECS cannot be easily applied in 3N  
systems.  We could not overcome 
problems arising from  
the permutation operator (see Section III).

\section{Three-body resonances}

Now we investigate 3-body resonances as solutions of 
configuration space Faddeev equations. 
They are often called Kowalski-Noyes equations 
and have the form~\cite{7}
\begin{eqnarray}
( H_0 + V - E ) \psi = -V~P~\psi
\label{eq16}
\end{eqnarray}
Here $H_0$ is the kinetic energy, V the two-body force, $\psi$ a 
Faddeev component and P the sum of two permutation operators 
such that the total, properly symmetrized 
 wave function  is $\Psi = (1+P)\psi$. 
Based on standard Jacobi variables $\overrightarrow x = \overrightarrow {r_2} 
- \overrightarrow {r_3}$, $\overrightarrow y = {1\over{3}} 
(\overrightarrow {x_1} 
- {1\over{ 2}} ( \overrightarrow {x_2} + 
\overrightarrow {x_3} ) )$ it is convenient 
to introduce polar coordinates $x \equiv \rho cos\theta$ and 
$y \equiv {\sqrt3 \over{ 2}} 
\rho sin\theta$. 
If we express the Faddeev component 
$\psi(\overrightarrow x,\overrightarrow y)$  in terms of 
 partial wave amplitudes $\vert \alpha >$ which 
specify angular momenta and isospin quantum numbers, then we obtain 
\begin{eqnarray}
 \psi(\overrightarrow x, \overrightarrow y) = \sum_{\alpha} 
 {{\phi_{\alpha}(x,y)}\over{xy}} 
 < \hat x \hat y \vert \alpha >,
\label{eq17}
\end{eqnarray}
and Eq.(\ref{eq16}) turns into 
a coupled set of partial-integrodifferential equations~\cite{8} 
\begin{eqnarray}
(\bigtriangleup_\alpha + {{m}\over{\hbar^2}}E ) \phi_{\alpha}(\rho,\theta) &-& 
 \sum_{\beta} V_{\alpha \beta}(\rho cos\theta ) \phi_{\beta}(\rho,\theta) = 
\sum_{\beta} V_{\alpha \beta}(\rho cos\theta ) \nonumber \\
&\times& \sum_{\gamma} 
\int_{\theta_-}^{\theta+} K_{\beta \gamma}(\theta,\theta') 
\phi_{\gamma}(\rho,\theta')d\theta'
\label{eq18}
\end{eqnarray}
with 
\begin{eqnarray}
\bigtriangleup_\alpha \equiv 
 {{\partial^2}\over{\partial \rho^2}} +  {{1}\over{\rho}}
{{\partial}\over{\partial \rho}} +  {{1}\over{\rho^2}}
{{\partial^2}\over{\partial \theta^2}} - 
{{l(l+1)}\over{\rho^2 cos^2\theta}} - 
{{\lambda(\lambda+1)}\over{\rho^2 sin^2\theta}}
\label{eq19}
\end{eqnarray}
and $\theta^- = \vert \theta - {{\pi}\over{3}} \vert $, 
$\theta^+ = \vert \vert \theta - {{\pi}\over{6}} \vert  - 
{{\pi}\over{2}} \vert$. 

The coupling 
results from the permutation operator P and from possible spin 
dependencies in V. 
The evaluation of the permutation operator matrix elements 
$< xy\alpha \vert P \vert x'y'\alpha' >$ and the resulting expression 
for the kernel $K_{\beta \gamma}(\theta,\theta')$ is 
presented in Appendix A. 

Similarily as in the 2-body case the resonance conditions are such 
that $\psi(x,y) \equiv {{\phi(x,y)}\over{xy}}$ has to be regular at the origin 
and purely outgoing in all channels. For the reduced amplitudes 
$\phi (\rho,\theta)$ this amounts to~\cite{8,14}
\begin{eqnarray}
\phi(0,\theta)=\phi(\rho,0)=\phi(\rho,{{\pi}\over{ 2}})=0
\label{eq20}
\end{eqnarray}
and for large distances 
\begin{eqnarray}
\phi(\rho,\theta) \rightarrow u(x)e^{iqy}f_0 + 
{  {e^{ik\rho}}  \over ({k{\rho})^{ 1\over {2} } } }   
( A(\theta) + 
{1\over{{k\rho}}}B(\theta) + \cdots )
\label{eq21}
\end{eqnarray}
The first term in Eq.~(\ref{eq21}) occurs only in channels $\alpha$,  
 where a two-body bound state $u(x)$  
exists.  The second part describes the behavior in the 3-body breakup 
channel. The corresponding momenta are 
\begin{eqnarray}
q &=& \sqrt{m(E-E_2)} \nonumber \\ 
k &=& \sqrt{mE}
\label{eq22}
\end{eqnarray}
where m is the nucleon 
mass and $E_2$ the two-body binding energy. The boundary 
conditions (\ref{eq21}) apply also to bound states with real 
eigenvalues E which are located below all two-body  thresholds $E_2$. For 
resonances E is complex and the eigenvalues 
E close to the real axis of the physical sheet are of interest. They 
either lie in second sheets which are accessible to the right of $E_2$-values 
(two-body fragmentation cuts) or in the lower half plane of a second sheet, 
which is accessible through the 3N-breakup cut to the right of $E=0$. 
For a recent description of the cut-structure see for instance~\cite{10}. 

An efficient method for solving Eq.~(\ref{eq18}) is very well displayed 
in~\cite{8}. One chooses suitable grid points in $\rho$ and $\theta$ 
and interpolates $\phi(\rho,\theta)$ in terms of cubic splines. 
Similarily as in the 2-body case the boundary conditions 
 (\ref{eq20}) and  (\ref{eq21}) can be conveniently incorporated 
either by the absence of certain expansion coefficients in the spline 
expansion or by suitable relations among the coefficients. Then choosing 
appropriate collocation points in $\rho$ and $\theta$ one ends up in a 
homogeneous algebraic set of equations. For realistic NN forces the dimension 
of the matrix is very large and the tensor product method~\cite{11} 
is a key-technique. We use a Lanczos type iteration method~\cite{13} 
and of course take advantage of the sparceness property of the matrices. 
It is advantegeous to work with further reduced amplitudes F defined as 
$\phi(\rho,\theta)\equiv 
\rho e^{ik\rho} F(\rho,\theta)$, if one treats resonances below the cut 
with $Re~E > 0$. 
In Appendix B we present the details of the above procedure.

The main problem in solving Eq.~(\ref{eq18}) is the approximate 
nature of the 
asymptotic form (\ref{eq21}). Ab initio it is not known at which $\rho$-value 
this form is valid with sufficient accuracy. For a recent discussion 
see~\cite{14}. 
Also it is a priori not known how 
many terms should be included in the asymptotic 
expansion (\ref{eq21}) for a particular 3N resonance state.  
 One suspects of course that including more terms one approaches more closely 
the exact boundary condition. 
The importance of the terms in Eq.(\ref{eq21}) going with 
increasing powers of ${{1}\over{k\rho}}$ depends not only on the 
resonance energy and the matching value for $\rho$ but also on the 
magnitudes of the a priori unknown functions $A(\theta), B(\theta), \cdots$. 
For resonances commonly these problems are expected  to be overcome using 
the complex scaling method (CSM)~\cite{15}. 
However, due to the 
restrictions imposed on this method by the behavior of the NN forces 
as discussed in Section II we found that this method  
can be  applied 
in the framework of the configuration space Faddeev equations 
only  for rather narrow 
resonances. 
Thereby all relative coordinates ( x and y in the 3-body case)  
are  rotated into the complex plane. This amounts to 
$\rho \rightarrow {\rho}'={\rho}e^{i\alpha}$ where  $\alpha$ is restricted 
to rather small values. An example for realistic nn forces 
is mentioned at the end of this Section. However, if the forces are 
very simple, CSM can work as will be illustrated below. 

 One is intended to 
hope that SECSM which proved to be very efficient in the 2-body 
case will help here, too. However, according to our insights up to now 
this approach leads to  problems we could not solve 
when calculating the permutation operator P and when performing 
the transition from cartesian to polar relative coordinates. 

The only possibility left  is 
to look at solutions of Eq.(\ref{eq18}) without leaving 
the 
real r-axis and taking the boundary condition (\ref{eq21}) with  a sufficient  
 number of terms into account in order to reach as close as possible the exact 
boundary condition. In this manner one can hope 
to get  convergence for the resulting resonance 
position. 

We checked our configuration space code 
by calculating the energy of the bound state of $^3H$ 
nuclei with the Reid 93 NN potential taking different NN force components 
into account. In Table~\ref{table3a} we present our numbers together 
with results of momentum space Faddeev calculations~\cite{noga99}. 
The very good agreement between both approaches is clearly seen. 

In the first step we considered 3 bosons interacting 
via Gaussian S-wave forces 
\begin{eqnarray}
V(r) = -120 exp(-r^2) + 12 exp(-r/3)^2 ~~~ [MeV].
\label{eq24}
\end{eqnarray}
They are such that a barrier is built in, which is an obvious and trivial 
mechanism 
generating a resonance. That interaction does not support a two-body bound 
state and in the asymptotic form (\ref{eq21}) the first term is absent. 
For 3 bosons in the 
$J^{\pi}=0^+$ state and interacting by  S-wave forces only the kernel 
$K_{\alpha \beta}={{1}\over{4\pi}}$. This model has been used in~\cite{17} 
in applying the stochastic variational method. 
The resonance energy found there is $E=(8.60 - i~1.84)$~MeV. 
In this case CSM works. 
For instance for a 
matching radius of $\rho=10$~fm and taking the A- and B-terms 
in (\ref{eq21}) into account we achieved the results shown in 
Table~\ref{table4}. For $Re(\alpha)=40.0^o$ we are  close 
to the result found in~\cite{17} but not close enough. 
A smaller matching radius, however, like 
  $\rho=5$~fm does not work. The correct result can however be achieved 
if in addition we turn on $Im(\alpha)<0$. This is demonstrated in 
 Table~\ref{table4}. 
We illustrate in Figs.\ref{fig2} and \ref{fig3} typical 
resonance amplitudes F  working with real coordinates or with 
complex coordinates. In both cases the matching radius 
has been chosen as  $5$~fm. In the 
first case this is not sufficient to achieve the correct complex 
energy eigenvalue, whereas in the second case the amplitude is already 
sufficiently damped at $\rho=5$~fm and the correct energy 
eigenvalue is accessible at such a small matching radius. 

In this simple case one can study the problem  
how many terms $A, B, \cdots$ in the boundary condition 
(\ref{eq21}) should be taken into account  to get the correct
 resonance position. 
Of course it will depend on the value of the matching radius $\rho$. 
We present in Table~\ref{table5} resonance positions obtained with 
increasing number of terms in the boundary condition (\ref{eq21}) 
and different matching radii. 
 For the details 
to include arbitrary number of terms in the asymptotic condition 
we refer to Appendix B. It is seen from 
Table~\ref{table5} that for matching radii $> 10$~fm it is practically 
sufficient  to take only the two leading terms going with A and B into 
account in order to 
get the correct resonance position. We also  see that the 
restriction to the A term alone  
is totally insufficient. Unfortunately this example is far from reality 
in nuclear physics. It does not contain a short range repulsion 
and the mechanism for creating a resonance in few nucleon systems is 
different in reality, where it is not caused by a barrier in the potential 
itself. 

Now we investigate the case of 
three interacting neutrons in  states with different 
total angular momenta and parity. We 
allowed the nn force to act in the states $^1S_0,~^3P_0$ and 
$^3P_2-^3F_2$. They are of attractive nature while the repulsive ones were 
neglected. This is only a feasibility study and should provide a first 
orientation. 
In the first step the nn forces were chosen as a smooth,  
local NN force of ref.~\cite{gogny}, called in the following Gogny 
potential. It is a rather simple approach to  NN phase-shifts 
through forces of central, tensor and spin-orbit type supplemented by a  
$L_{12}$ term, all of which of Gaussian form. 
To initiate the search 
process for the position of the resonance energy we artificially enhanced 
the force components in the three states such that all of 
them support a two-neutron 
bound state at the same energy. An example is shown in Table~\ref{table6} 
for a  $J^{\pi} = 3/2^-$ state of three neutrons. 
For these enhancement factors 
also the 3n system is bound. Then we reduce the enhancement 
factors (some intermediate cases are shown in Table~\ref{table6}), 
which traces out a bound state trajectory. In this case we see that 
the  3n bound state still exists 
even when all 2n subsystem bound states have disappeared. Then 
we can neglect the first term in the boundary condition  (\ref{eq21}).  
Decreasing further the enhancement factor(s) (in this specific case 
only for $^3P_2-^3F_2$) the energy eigenvalue goes into the second sheet and 
traces out a resonance trajectory. 
The search for the resonance position corresponding to the actual nn force 
is finished when the 
enhancement factors in all nn force components have reached 
 the 
values $\lambda_i=1.0$. As we shall see in the case of  three neutrons 
 we face numerical instabilities before all $\lambda's$ reach the value 1. 
We can present the resonance trajectory
only partially up to the point, where we encounter those instabilities. 
We expect that they are probably caused by the difficulty 
to fulfill the resonance condition, which requires that the 
exponentially decreasing, purely incoming wave is much smaller 
than the exponentially increasing purely outgoing wave. 
As we see from Table~\ref{table6} the $J^{\pi} = 3/2^-$ 3n 
bound state disappears when the enhancement factors arrive at 
$(\lambda_{^1S_0}, \lambda_{^3P_0}, \lambda_{^3P_2-^3F_2})=(1.0, 1.0, 4.35)$. 
 In Table~\ref{table7} the trajectory of the 
$3/2^-$ 3n resonance in the second 
energy sheet is mapped out by decreasing the enhancement factor 
$\lambda_{^3P_2-^3F_2}$ in step of 0.05 starting from 
$\lambda_{^3P_2-^3F_2}=4.3$. The final reliable position reached is for 
$\lambda_{^3P_2-^3F_2}=3.6$ with the value 
$E_{3n}^{3/2^-} = (5.30 -i4.67)$~MeV. 
As can be seen this stable position is only obtained 
when at least 4 terms in the boundary condition (\ref{eq21}) are taken into 
account. We checked that our particular distribution of $\rho$ and $\theta$ 
points as well as the choice of the matching radius in $\rho$ does not 
influence the resulting trajectory. The values given 
in Table~\ref{table7} are based on a matching radius $\rho=30.0$~fm  and 
typically 
$N_{\rho}=30$ and $N_{\theta}=25$ points distributed by 
scaling factors  $S_{\rho}=1.2$ and $S_{\theta}=0.9$ (see ref.~\cite{8}). 
It is unfortunate that right now we are not able to 
reach the final position with ${\lambda}_{^3P_2-^3F_2}=1$. 
However very likely 
the width will be quite large. 
Therefore one cannot expect to see a visible enhancement around 
some 3n c.m. energy in a reaction producing three neutrons in the 
state $3/2^-$. 

Similarily to $3/2^-$ behave the $1/2^-$ and  $3/2^+$  states based on the  
 Gogny potential. 
With proper enhancement factors $\lambda_V$ the three neutrons are bound 
 together with  2-neutron bound states for $^1S_0$, $^3P_0$ and $^3P_2-^3F_2$.
 These 3n states remain 
bound even when the 2-neutron  bound states disappear with appropriately 
reduced enhancement factors.
They disappear into  the second energy sheet 
 when the enhancement factors reach 
$(1.0, 7.5, 4.75)$ and $(1.0, 5.5, 4.8)$ 
for $J^{\pi} = 1/2^-$ and $3/2^+$, respectively. 
These enhancements factors are of course not unique but we worked with 
those examples. 
Following their trajectories 
before numerical instabilities set in 
results in the complex energy eigenvalues 
 shown for the Gogny potential in 
Table~\ref{table8}. 
For the $1/2^-$ and  $3/2^+$ states, similarily as for 
 $3/2^-$, the final, converged position is obtained only when 
at least 4 terms in the boundary condition are taken into account. 
In Fig.\ref{fig4} we show the trajectories for the three states 
$1/2^-$, $3/2^-$, and  $3/2^+$ for the Gogny potential. We note that the 
trajectories are not unique and depend how the enhancement factors 
are changed. Our startegy was to bring all of them together to one 
as close as possible. 

 In this article we did not investigate the state 
$1/2^+$, since we found, like 
in  ~\cite{glo78} that the 3n bound state disappears earlier than the 2n 
bound states. In other words, the 3n bound state trajectory will enter into 
the 2n-n two-body fragmentation cut. We left that study with the 
additional asymptotic term  in Eq.(\ref{eq21}) to a future investigation. 

It is seen from Table~\ref{table8} that the 
$3/2^+$ state for the Gogny potential is 
relatively narrow. Taking the enhancement factors $\lambda_i=(1.0,5.3,4.6)$ 
results in a resonance position $E_{3n}^{3/2^+} = (2.88 -i0.23)$~MeV. 
For such a narrow resonance the CSM works properly for not too large 
values of the $\alpha$ angle (see Table~\ref{table8a}). 
Increasing it, however, the resulting energy eigenvalue is totally 
wrong what can be traced back to the increasing oscillations of the NN 
potential resulting from its analytical continuation to the complex 
r-values.

In the last step we checked how the results depend on the NN interaction used. 
We choose a  modern, realistic NN interaction as given by 
the Reid 93 potential. 
The resulting positions of the 3n resonances, as far as we could trace 
them out,  are shown for two examples 
in Table~\ref{table9}. In spite of the fact that the 
particular 
values of the $\lambda_V$'s change of course, the resulting picture is 
basically the same as for the Gogny potential. 
We are still too far away from all $\lambda_i=1$ in order to definitely 
say that all the widths will be quite large. But this will likely 
be the case. Again to achieve converged results one has to 
take at least 4 terms in the boundary condition (\ref{eq21}) into account. 

\section{Summary}

We investigated the configuration space Faddeev equations with the 
aim to determine the location of 3-neutron resonances. 
Realistic nn forces have been used. The asymptotic form 
in the 3n breakup channel is known only in the sense of an 
asymptotic expansion. We found that several terms thereof 
have to be included in order to get stable and reliable 
results, which are independent of the matching radius. 
The search for the resonance positions was carried through by 
artificially enhancing 2n force components such that a 3n bound 
state exists. Then by reducing those enhancement factors 
 we followed the paths of the energy eigenvalue into the second sheet 
adjacent to $Re(E) \ge 0$. Unfortunately it turned out that numerical 
instabilities set in before we reached the actual strengths factors 1 
for all the individually enhanced force components. 
Therefore we could not reach the very final positions of the studied 
3n resonances in the states $J^{\pi}=1/2^-, 3/2^-$, and  $3/2^+$. 
The detailed reason for those instabilities remains to be found out. 
Despite of that our results already indicate that the imaginary parts 
of the complex resonance energies will be quite large, so that one should not 
expect to see resonances in experiments. 

We also applied the CSM, which works beautifully in atomic phyics. 
The complex coordinates turn the exponentially increasing resonance 
wave function 
into an exponentially decreasing one, like for a bound state, and 
the problem with the approximate boundary conditions is avoided.  
However, for realistic NN interactions the complex scaling transformation 
$r  \to  r~e^{i\alpha}$ increases the interaction range 
by a factor ${{1}\over{cos\alpha}}$ and leads for scaling angles $\alpha$ 
greater than about $40^o$  to a strong oscillations of the potential. 
This restricts the application of the complex scaling method in 
more than 2-nucleon systems to  rather narrow resonances. 
In the 2-nucleon system one can avoid the above restriction 
by performing the complex rotation outside the region of non-zero 
interaction. 
Due to the neccessary particle permutation in the 3-body system 
we were not able to use that idea in the 3-body context. 

This project deserves further investigations to locate 
all lowest resonances in the three nucleon systems based on modern 
NN and possibly three-nucleon forces.

\begin{center}
{\bf Acknowledgements}
\end{center}

This work was supported by the Deutsche Forschungsgemeinschaft 
 under Project No. Gl87/24-1. 
 The numerical calculations have been performed on the 
CRAY T90 and the CRAY T3E of the H\"ochstleistungsrechenzentrum in J\"ulich,
Germany.

\pagebreak


\appendix
\section{Permutation operator}

To evaluate the term on the right hand side of Eq.(\ref{eq16}) one needs 
\begin{eqnarray}
< x y \beta \vert  P \vert \psi > = \sum_{\gamma} \int dx' x'^2 dy' y'^2 
< x y \beta \vert P \vert x' y' \gamma > {{\phi_{\gamma}(x',y')}\over{x'y'}}
\label{a1}
\end{eqnarray}
where the  permutation operator is $P \equiv P_{12}P_{23} + P_{13}P_{23}$ and 
the set of  
discrete quantum numbers abbreviated by 
$\beta$ or $\gamma$ is $\lbrace (l s)j (\lambda 1/2)I 
 (j I)JM; (t 1/2)TM_T \rbrace$. Here l,s, and j are the 
orbital angular momentum, 
total spin and total angular momentum in the two-body subsystem 2-3, 
$\lambda$, 1/2, and I are the corresponding quantities for particle 1, and JM 
denote the conserved 3N total angular momentum and its magnetic quantum 
number. In isospin space, t refers to particles 2 and 3 and couples 
with the isospin 1/2 of particle 1 to the 
total isospin T and its magnetic quantum number $M_T$. 
Following~\cite{gl83} one obtains  for the matrix element of the permutation 
operator 
\begin{eqnarray}
< x y \alpha \vert  P \vert x' y' \alpha' > = \int_{-1}^{1} du 
{{\delta(x'- \tilde x')}\over{(x')^{l'+2}}} 
{{\delta(y'- \tilde y')}\over{(y')^{\lambda'+2}}} 
\tilde G^{cart}_{\alpha \alpha'}(x,y,u) 
\label{a2}
\end{eqnarray}
with 
\begin{eqnarray}
\tilde x' &=& \sqrt{ {{1}\over{4}} x^2 + x y u + y^2} \nonumber \\ 
\tilde y' &=& \sqrt{ {{9}\over{16}} x^2 - {{3}\over{4}} x y u + 
{{1}\over{4}} y^2}
\label{a3}
\end{eqnarray}
and 
\begin{eqnarray}
\tilde G^{cart}_{\alpha \alpha'}(x,y,u) = \sum_k P_k(u) 
\sum_{r_1+r_2=l'} \sum_{s_1+s_2=\lambda'} (x)^{r_1 + s_1} 
(y)^{r_2 + s_2} \tilde g_{\alpha \alpha'}^{r_1s_1k}  
\label{a4}
\end{eqnarray}
The purely geometrical quantity  $\tilde g_{\alpha \alpha'}^{r_1s_1k}$ 
is given by 
\begin{eqnarray}
\tilde g_{\alpha \alpha'}^{r_1s_1k} &=& -\sqrt{\hat l \hat l' \hat s 
\hat s' \hat j \hat j' \hat {\lambda} \hat {\lambda}'  \hat I \hat I'  
\hat t \hat t' } (-1)^{k + l' + s_1} \hat k 
({{1}\over{2}})^{r_1 + s_2}  ({{3}\over{4}})^{s_1} \nonumber \\
&&\sum_{LS} (-1)^L \hat L \hat S 
{\left\lbrace\matrix{ l & s & j \cr
                     \lambda & {{1}\over{2}} & I \cr
                     L & S & J \cr } \right\rbrace } 
{\left\lbrace\matrix{ l' & s' & j' \cr
                     {\lambda}' & {{1}\over{2}} & I' \cr
                     L & S & J \cr } \right\rbrace }
{\left\lbrace\matrix{ 1/2  & 1/2  & t \cr
                      1/2  & T    & t' \cr } \right\rbrace } 
{\left\lbrace\matrix{ 1/2  & 1/2  & s \cr
                      1/2  & S    & s' \cr } \right\rbrace } \nonumber \\ 
&&\sqrt{ {{ \hat l'! }\over{(2r_1)!(2r_2)!  }}  }  
\sum_{r_3} \hat r_3 {\left(\matrix{ r_1  & s_1  & r_3 \cr
                      0  & 0    & 0 \cr } \right) }  
 {\left(\matrix{ l  & r_3  & k \cr
                      0  & 0    & 0 \cr } \right) } \nonumber \\ 
&&\sqrt{ {{ \hat {\lambda}'! }\over{(2s_1)!(2s_2)!  }}  }   
\sum_{s_3} \hat s_3 {\left(\matrix{ r_2  & s_2  & s_3 \cr
                      0  & 0    & 0 \cr } \right) }  
 {\left(\matrix{ \lambda  & s_3  & k \cr
                      0  & 0    & 0 \cr } \right) }  \nonumber \\ 
&&{\left\lbrace\matrix{ l  & r_3  & k \cr
                      s_3  & \lambda    & L \cr } \right\rbrace }  
{\left\lbrace\matrix{ r_1 & r_2 & l' \cr
                      s_1 & s_2 & {\lambda}' \cr
                      r_3 & s_3 & L \cr } \right\rbrace } 
\label{a5}
\end{eqnarray}
The superscript $cart$ stands for cartesian and $\hat l \equiv 2l + 1$. 
To arrive at the kernel $K_{\beta \gamma}$ on 
the right-hand side of Eq.(\ref{eq18}) 
one has  to evaluate
\begin{eqnarray}
x y < x y \beta \vert  P \vert \psi > = \sum_{\gamma} x y \int_{-1}^{1} du 
{{\tilde G_{\beta \gamma}^{cart}(x,y,u)}\over{\tilde x'^{l_{\gamma}} 
\tilde y'^{{\lambda}_{\gamma}}  }} 
{{\phi_{\gamma}(\tilde x',\tilde y')}\over{\tilde x',\tilde y'}} 
\label{a6}
\end{eqnarray}
We introduce polar coordinates 
\begin{eqnarray}
\tilde x' &\equiv& \rho cos {\theta}' \nonumber \\
\tilde y' &\equiv& {{\sqrt{3}}\over{2}} \rho sin {\theta}'
\label{a7}
\end{eqnarray}
and changing the  $u$ to the $\theta'$ integration 
\begin{eqnarray}
\int_{-1}^1 d u {{xy}\over{\tilde x' \tilde y'}} \cdots = 
{{4}\over{\sqrt{3}}} \int_{{\theta}^-}^{{\theta}^+} d \theta'  \cdots
\label{a8}
\end{eqnarray}
one arrives at 
\begin{eqnarray}
x y < x y \beta \vert  P \vert \psi > = \sum_{\gamma} 
{{4}\over{\sqrt{3}}}
 \int_{{\theta}^-}^{{\theta}^+} d\theta' 
{{\tilde G_{\beta \gamma}^{polar}(\theta,\theta')}\over{
(cos\theta')^{l_{\gamma}} 
({{\sqrt{3}}\over{2}}sin\theta')^{{\lambda}_{\gamma}}  }} 
\phi_{\gamma}(\rho,\theta') 
\label{a9}
\end{eqnarray}
with
\begin{eqnarray}
\tilde G^{polar}_{\beta \gamma}(\theta,\theta') = 
\sum_k P_k(u(\theta,\theta')) 
\sum_{r_1+r_2=l_{\gamma}} \sum_{s_1+s_2={\lambda}_{\gamma}} 
(cos\theta)^{r_1 + s_1} 
({{\sqrt{3}}\over{2}}sin\theta)^{r_2 + s_2} 
\tilde g_{\beta \gamma}^{r_1s_1k}  
\label{a10}
\end{eqnarray}
and 
\begin{eqnarray}
u(\theta,\theta') = {{ cos^2\theta' - {{1}\over{4}}cos^2\theta 
 - {{3}\over{4}}sin^2\theta }\over{ {{\sqrt{3}}\over{2}}cos\theta sin\theta }}
\label{a11}
\end{eqnarray}
From Eq.(\ref{a9}) follows that
\begin{eqnarray}
K_{\beta \gamma}(\theta,\theta') = 
{{4}\over{\sqrt{3}}} 
{{\tilde G_{\beta \gamma}^{polar}(\theta,\theta')}\over{
(cos\theta')^{l_{\gamma}} 
({{\sqrt{3}}\over{2}}sin\theta')^{{\lambda}_{\gamma}}  }} 
\label{a12}
\end{eqnarray}

\pagebreak

\section{The Solution of the energy eigenvalue problem 
including n-terms in the asymptotic 
boundary condition}

Due to numerical reasons it is advantegous to extract the exponential part 
from the reduced Faddeev amplitude $\phi_{\alpha}(\rho,\theta)$ 
and to work instead with 
\begin{eqnarray}
 F_{\alpha}(\rho,\theta) \equiv \phi_{\alpha}(\rho,\theta) 
/ (\rho e^{ik\rho})
\label{b1}
\end{eqnarray}
Following ref.~\cite{8} one introduces suitable grid points in 
$\rho~~\lbrace\rho_i,~~ i=0, \cdots,I$;~~ 
$\rho_0=0, \rho_I=\rho_{max} \rbrace$ and 
in $\theta~~\lbrace \theta_k,~~ k=0, \cdots,K;~~ \theta_0=0, 
\theta_K={{\pi}\over{2}} \rbrace$.  
Then one can write $F_{\alpha}(\rho,\theta)$ in terms of cubic splines 
$s_m(\rho)$ 
and $s_n(\theta)$ as 
\begin{eqnarray}
 F_{\alpha}(\rho,\theta) = 
\sum_{m=0}^{M+1} \sum_{n=0}^{N+1} a_{\alpha m n}  
  s_m(\rho) s_n(\theta)
\label{b2}
\end{eqnarray}
with $M=2I$ and $N=2K$.

Due to the properties of the cubic splines as given in~\cite{8} one has
\begin{eqnarray}
 F_{\alpha}(\rho_i,\theta_k) &=&  a_{\alpha,2i,2k} \nonumber \\
 {{\partial F_{\alpha}}\over{\partial \rho}}\vert_{\rho_i,\theta_k} 
&=&  a_{\alpha,2i+1,2k} \nonumber \\
 {{\partial F_{\alpha}}\over{\partial \theta}}\vert_{\rho_i,\theta_k} 
&=&  a_{\alpha,2i,2k+1}
\label{b3}
\end{eqnarray} 
The boundary conditions at the origin (\ref{eq20}) imply that
\begin{eqnarray}
a_{\alpha,0,n}  &=& 0  \nonumber \\
a_{\alpha,m,0}  &=& 0  \nonumber \\
a_{\alpha,m,n=N}  &=& 0.
\label{b4}
\end{eqnarray}
Writing the asymptotic expansion (\ref{eq21}) for $F_{\alpha}(\rho,\theta)$ 
under the assumption that no two-body bound state exists in any 
channel $\alpha$ results in
\begin{eqnarray}
 F_{\alpha}(\rho,\theta) \to  {{1}\over{\rho(k\rho)^{1/2}}} 
\sum_{l=1}^{N_t} {{B_l^{\alpha}(\theta)}\over{(k\rho)^{l-1}}}
\label{b5}
\end{eqnarray} 
when $N_t$ terms are included in the asymptotic form.

Imposing the 
boundary condition (\ref{b5}) at $\rho = {\rho}_{max}$  implies  
the dependence of  $a_{\alpha,m=M+1,n}$ on  $a_{\alpha,m=M,n}$ and 
 $B_l^{\alpha}(\theta)$. In order to remove the unknown $B_l^{\alpha}(\theta)$ 
functions the boundary condition has to be imposed on the last $N_t$ points 
of the $\rho$-grid. This leads to the dependence of 
$a_{\alpha,m=M+1,n}$ on $a_{\alpha,m=M,n}$, $\cdots$, 
$a_{\alpha,m=M-2(N_t - 1),n}$. In the following the form of this 
dependence is derived. 

The asymtotic form for the partial derivative  
${{\partial F_{\alpha}}\over{\partial \rho}}$ is according to (\ref{b5}) 
\begin{eqnarray}
 {{\partial F_{\alpha}(\rho,\theta)}\over{\partial \rho}} \to 
-\tilde A_1(\rho) F_{\alpha}(\rho,\theta) - \sum_{l=2}^{N_t} 
\tilde A_l(\rho) B_l^{\alpha}(\theta)
\label{b6}
\end{eqnarray} 
with 
\begin{eqnarray}
\tilde A_1(\rho) =  {{3}\over{2\rho}}  
\label{b7}
\end{eqnarray} 
and for $l \ge 2$
\begin{eqnarray}
\tilde A_l(\rho) =  {{l-1}\over{{\rho}^2(k_0\rho)^{l-1/2}}}.
\label{b8}
\end{eqnarray} 
Taking the partial derivative of (\ref{b2}) with respect to $\rho$ 
and the form  (\ref{b6}) 
yields for the last $N_t$  points of the  $\rho$ grid 
$~~\lbrace \rho = \rho_{I-i+1},~~ i=1, \cdots, N_t \rbrace$ 
\begin{eqnarray}
\sum_n a_{\alpha,M+1-2(i-1),n} s_n(\theta)  = 
- \sum_n a_{\alpha,M-2(i-1),n} s_n(\theta) \tilde A_1(\rho_{I-i+1}) 
- \sum_{l=2}^{N_t} \tilde A_l(\rho_{I-i+1}) B_l^{\alpha}(\theta).
\label{b9}
\end{eqnarray}
Eq.(\ref{b9}) together with 
its $\theta$-derivative taken at the points of the  
$\theta$-grid lead to 
\begin{eqnarray}
a_{\alpha,M+1-2(i-1),n}  = -a_{\alpha,M-2(i-1),n} \tilde A_1(\rho_{I-i+1}) 
- \sum_{l=2}^{N_t} \tilde A_l(\rho_{I-i+1}) \tilde B_l^{\alpha}(\theta_k)
\label{b10}
\end{eqnarray}
where for $n=2k$~~~~$\tilde B_l^{\alpha} = B_l^{\alpha}(\theta_k)$ and for 
 $n=2k+1$~~~~
$\tilde B_l^{\alpha} 
= {{d B_l^{\alpha}}\over{d \theta}}{\vert}_{\theta=\theta_k}$. 

The set (\ref{b10}) 
are  $N_t$ equations for $N_t - 1$ unknowns  
$\tilde B_l^{\alpha}(\theta_k)$. They allow to 
 write $a_{\alpha,M+1,n}$  in terms 
of  the $a_{\alpha,M-i+1,n}$, $i=1, \cdots, 2(N_t-1) + 1$ as 
\begin{eqnarray}
a_{\alpha,M+1,n}  = \sum_{i=1}^{2(N_t-1)+1} a_{\alpha,M-i+1,n} C_i
\label{b11}
\end{eqnarray}
where the $C_i$'s are functions of all the $\tilde A_l(\rho_{I-i+1})$'s 
 resulting 
from (\ref{b10}). 

In this way, incorporating the boundary conditions (\ref{b4}) and (\ref{b5}) 
one can write $F_{\alpha}(\rho,\theta)$ on the chosen grids as 
\begin{eqnarray}
 F_{\alpha}(\rho,\theta) = 
\sum_{m=1}^{M} \sum_{n=1}^{N} a_{\alpha m n}  s_m(\rho) s_n(\theta) + 
\sum_{n=1}^{N}\lbrack \sum_{i=1}^{2(N_t-1)+1} a_{\alpha,M-i+1,n} C_i \rbrack 
  s_{M+1}(\rho) s_n(\theta)
\label{b12}
\end{eqnarray}
where, due to the last equation in (\ref{b4}) we renumbered 
$n=N+1$ by $n=N$. 

Inserting the expansion 
(\ref{b12}) with the help of (\ref{b1}) 
into Eq.(\ref{eq18}) and choosing collocation 
points $\rho_p,~~ p=1, \cdots, 2I$ and $\theta_q,~~ q=1, \cdots, 2K$~\cite{8} 
yields  
\begin{eqnarray}
 \sum_{\beta =1}^{N_c} \sum_{m=1}^{M} \sum_{n=1}^{N} A_{\alpha p q,\beta m n} 
  a_{\beta m n} = 
 \sum_{\beta =1}^{N_c} \sum_{m=1}^{M} \sum_{n=1}^{N} B_{\alpha p q,\beta m n} 
  a_{\beta m n}
\label{b13}
\end{eqnarray}
with
\begin{eqnarray}
 A_{\alpha p q,\beta m n} &=& \delta_{\alpha \beta} 
\lbrack  
 s_{m}^{''}(\rho_p) s_n(\theta_q) + 
 ( {{3}\over{\rho_p}} + 2ik ) s_{m}^{'}(\rho_p) s_n(\theta_q) +
{{1}\over{\rho_p^2}} s_{m}(\rho_p) s_n^{''}(\theta_q) \nonumber \\
&&- 
\lbrace 
 {{l(l+1)}\over{\rho_p^2 cos^2\theta_q}} + 
 {{\lambda(\lambda+1)}\over{\rho_p^2 sin^2\theta_q}}
 -  ( {{1}\over{\rho_p^2}} + {{3ik}\over{\rho_p}} )
\rbrace s_{m}(\rho_p) s_n(\theta_q)
\rbrack 
\label{b14}
\end{eqnarray}
and
\begin{eqnarray}
 B_{\alpha p q,\beta m n} &=& V_{\alpha \beta}(\rho_p cos\theta_q) 
 s_{m}(\rho_p) s_n(\theta_q) \nonumber \\ 
&& + \sum_{\gamma=1}^{N_c} V_{\alpha \gamma}(\rho_p cos\theta_q) 
  s_{m}(\rho_p) \int_{\theta^-}^{\theta^+} K_{\gamma \beta}(\theta_q,\theta') 
 s_n(\theta')d \theta' \nonumber \\
&& + \delta_{m,M-i+1~(1\le i \le 2(N_t - 1)+1 )}
\Bigl \lbrace 
 -\delta_{\alpha \beta} C_i  
\lbrack  
 s_{M+1}^{''}(\rho_p) s_n(\theta_q)  \nonumber \\ 
&& + ( {{3}\over{\rho_p}} + 2ik ) s_{M+1}^{'}(\rho_p) s_n(\theta_q) +
{{1}\over{\rho_p^2}} s_{M+1}(\rho_p) s_n^{''}(\theta_q) \nonumber \\
&&- 
\lbrace 
 {{l(l+1)}\over{\rho_p^2 cos^2\theta_q}} + 
 {{\lambda(\lambda+1)}\over{\rho_p^2 sin^2\theta_q}}
 -  ( {{1}\over{\rho_p^2}} + {{3ik}\over{\rho_p}} )
\rbrace s_{M+1}(\rho_p) s_n(\theta_q)
\rbrack \nonumber \\
&& V_{\alpha \beta}(\rho_p cos\theta_q) 
 s_{M+1}(\rho_p) s_n(\theta_q) \nonumber \\ 
&& + \sum_{\gamma=1}^{N_c} V_{\alpha \gamma}(\rho_p cos\theta_q) 
  s_{M+1}(\rho_p) \int_{\theta^-}^{\theta^+} 
K_{\gamma \beta}(\theta_q,\theta') 
 s_n(\theta')d \theta' 
\Bigr \rbrace
\label{b15}
\end{eqnarray}
The number of channels is denoted by $N_c$. 
This eigenvalue problem (\ref{b13}) can be treated in a 
standard way by changing it into the form 
\begin{eqnarray}
 A^{-1} B a = \lambda a 
\label{b16}
\end{eqnarray}
and searching  for that energy at which $\lambda=1$.

To solve Eq.(\ref{b16}) we used a Lanczos type iteration method~\cite{13}. 
For the application of B and $A^{-1}$ on a given vector a we applied  
the tensor 
product method~\cite{11}. Hereby  A is decomposed as 
\begin{eqnarray}
 A = \bar A \otimes \bar B + \bar C \otimes \bar D
\label{b17}
\end{eqnarray}
with
\begin{eqnarray}
\bar A_{p m} &\equiv&  s_{m}(\rho_p)
 ( {{1}\over{\rho_p^2}} + {{3ik}\over{\rho_p}} ) + 
({{3}\over{\rho_p}} + 2ik ) s_{m}^{'}(\rho_p) +s_{m}^{''}(\rho_p) \nonumber  \\
\bar B_{q n}^{\alpha \beta} &\equiv&  \delta_{\alpha \beta} 
 s_n(\theta_q) \nonumber  \\ 
\bar C_{p m} &\equiv&  {{1}\over{\rho_p^2}} s_{m}(\rho_p) \nonumber  \\
\bar D_{q n}^{\alpha \beta} &\equiv&  \delta_{\alpha \beta} 
\lbrack 
 s_n^{''}(\theta_q) - 
( 
 {{l(l+1)}\over{cos^2\theta_q}} + 
 {{\lambda(\lambda+1)}\over{sin^2\theta_q}} 
) s_n(\theta_q)
\rbrack .
\label{b18}
\end{eqnarray}
Then we use the identity 
\begin{eqnarray}
  \bar A \otimes \bar B + \bar C \otimes \bar D = 
(\bar A U^{-1} \otimes \bar B V^{-1}) (\Pi \otimes \Xi + I \otimes I) 
(U \otimes V)
\label{b19}
\end{eqnarray}
with 
\begin{eqnarray}
 \bar A^{-1} \bar C = U^{-1} \Pi U \nonumber  \\
 \bar B^{-1} \bar D = V^{-1} \Xi V
\label{b20}
\end{eqnarray}
and $\Pi, \Xi$ diagonal.

It follows 
\begin{eqnarray}
  (\bar A \otimes \bar B + \bar C \otimes \bar D)^{-1} = 
(U^{-1} \otimes V^{-1})  (\Pi \otimes \Xi + I \otimes I)^{-1} 
(U \bar A^{-1} \otimes V \bar B^{-1}) .
\label{b21}
\end{eqnarray}


\pagebreak

\begin{table}

\begin{tabular} {|c|c|c|c|c|c|c|}
 NN state&${\lambda}_V$&$r_{max}$&\multicolumn{4}{c|}{ $E^{2N}_{res}$ (MeV)  }\\
\hline
    & & &exact (m=l)&m=0&m=1&m=2\\
\hline
   $^3P_0$&5.50&10.0&1.42-i0.75&1.75-i1.04& &  \\
          &    &15.0&1.42-i0.75&1.58-i0.63& &  \\
          &    &20.0&1.42-i0.75&1.36-i0.58& &  \\
\hline
   $^1D_2$&5.00&10.0&15.08-i0.80&15.11-i0.90&15.08-i0.80&  \\
          &    &15.0&15.08-i0.80&15.07-i0.85&15.08-i0.80&  \\
          &    &20.0&15.08-i0.80&15.06-i0.83&15.08-i0.80&  \\
\hline
 $^3P_2-^3F_2$&6.20&10.0&2.79-i1.17&3.07-i1.05&2.79-i1.17&  \\
              &    &15.0&2.79-i1.17&2.62-i0.99&2.79-i1.17&  \\
              &    &20.0&2.79-i1.17&3.08-i1.28&2.79-i1.17&  \\
\end{tabular}
\caption{\label{table1} Resonance positions in different NN states 
obtained with the correct (column 4) and approximate (columns 5-7) 
boundary conditions  applied at different 
matching radii $r_{max}$. The Reid93 NN potential multiplied with 
the factors ${\lambda}_V$ has been used.}

\newpage

\begin{tabular}{|c|c|c|c|} 
r-matching & Re($\alpha$) & Im($\alpha$) & $E^{2N}_{res}$  \\
(fm)       &  ($^o$)    &  ($^o$)    &  (MeV)     \\  \hline
10.0       &  0.0       &  0.0       &  1.75 - i~1.04 \\
10.0       &  20.0      &  0.0       &  1.40 - i~0.88 \\
10.0       &  0.0       &  -40.0     &  1.36 - i~0.58 \\
10.0       &  0.0       &  -80.0     &  1.24 - i~0.58 \\
10.0       &  20.0      &  -40.0     &  1.42 - i~0.73  \\
10.0       &  20.0      &  -80.0     &  1.42 - i~0.75  \\
2.0        &  20.0      &  -80.0     &  1.22 - i~0.93   \\
2.0        &  20.0      &  -160.0    &  1.43 - i~0.75 \\
20.0       &  20.0      &  0.0       &  1.42 - i~0.73 \\
20.0       &  30.0      &  0.0       &  1.43 - i~0.74  \\
20.0       &  40.0      &  0.0       &  1.43 - i~0.75  \\
20.0       &  44.0      &  0.0       &  1.48 - i~0.75  \\
20.0       &  46.0      &  0.0       &  1.71 - i~0.72  \\
20.0       &  48.0      &  0.0       &  2.99 - i~0.36  \\
\end{tabular}
\caption{Complex energy eigenvalues for the $^3P_0$ 2N state obtained with the 
Reid~93 
NN interaction enhanced by $\lambda_V = 5.5$ and an approximate 
boundary condition ($m=0$ term only in Eq.(\ref{eq8})~). 
The correct energy eigenvalue 
is $E^{2N}_{res}=(1.42-i~0.75)$~MeV.}\label{table2}

\newpage

\begin{tabular}{|c|c|c|c|c|} 
r-matching &$\lambda$ &$r_0$  &$\alpha$&$E^{2N}_{res}$  \\
(fm)       & $(fm^-1)$  & $(fm)$     &  ($^o$)  &  (MeV)     \\  \hline
10.0&2.0&5.0&10.0&15.08 - i~0.84 \\
10.0&2.0&5.0&20.0&15.07 - i~0.81 \\
10.0&2.0&5.0&30.0&15.07 - i~0.80 \\
10.0&2.0&5.0&40.0&15.08 - i~0.80 \\
10.0&2.0&5.0&50.0&15.08 - i~0.80 \\
10.0&2.0&5.0&60.0&15.08 - i~0.80 \\
10.0&2.0&5.0&70.0&15.08 - i~0.80 \\
10.0&2.0&5.0&80.0&15.08 - i~0.80 \\
15.0&2.0&5.0&60.0&15.08 - i~0.80 \\
20.0&2.0&5.0&60.0&15.08 - i~0.80 \\
15.0&2.0&3.0&60.0&15.08 - i~0.80 \\
15.0&3.0&3.0&60.0&15.08 - i~0.80 \\
\end{tabular}
\caption{Smooth-exterior complex scaling applied to the $^1D_2$ resonance 
state of the Reid93 NN potential with $\lambda_V = 5.0$. Only the first 
term in 
the boundary condition (\ref{eq8}) was taken into account. 
The correct energy  
eigenvalue is $E^{2N}_{res}=(15.08-i~0.80)$~MeV.}\label{table3}

\newpage

\begin{tabular} {|c|c|c|c|}
 NN states&$N_c$&\multicolumn{2}{c|}{ $^3H$ binding energy (MeV)  }\\
\hline
    & &p-space &r-space\\
\hline
   $^1S_0 + ^3S_1-^3D_1$  &  5  & -7.630 & -7.631  \\
\hline
 $j \le 1$     &  10 & -7.403 & -7.406    \\
\hline
 $j \le 2$     &  18 & -7.747 & -7.750    \\
\end{tabular}
\caption{\label{table3a} Triton binding energies obtained 
in momentum- and coordinate-space 
 calculations with the Reid 93 NN potential using 
different numbers of channels $N_c$.}  

\newpage

\begin{tabular}{|c|c|c|c|}  
$\rho$-matching & Re($\alpha$) & Im($\alpha$) & $E^{3B}_{res}$  \\
(fm)       &    ($^o$)  & ($^o$)     &  (MeV)     \\  \hline
10.0       &  0.0       &  0.0       &  8.50 - i~1.81 \\
10.0       &  20.0      &  0.0       &  8.48 - i~1.87 \\
10.0       &  40.0      &  0.0       &  8.46 - i~1.86 \\
5.0        &  20.0      &  0.0       &  8.36 - i~1.78 \\
5.0        &  40.0      &  0.0       &  8.71 - i~2.03  \\
5.0        &  40.0      &  -20.0     &  8.45 - i~1.93  \\
5.0        &  40.0      &  -40.0     &  8.46 - i~1.86   \\
5.0        &  40.0      &  -60.0     &  8.47 - i~1.86   \\ 
\end{tabular}
\caption{CSM applied to a 3-boson system interacting through the S-wave
 interaction of Eq.(\ref{eq24}) in the $J^{\pi}=0^+$ state.  
 The correct energy eigenvalue 
is $E^{3B}_{res}=(8.47-i~1.86)$~MeV. The boundary condition (\ref{eq21}) 
with A and B terms only has been applied.}\label{table4}

\pagebreak

\begin{tabular} {|c|c|c|c|c|}
 $\rho_{max}$ (fm)&\multicolumn{4}{c|}{$E^{3B}_{res}~~(MeV)$ 
for particular number of terms in Eq.(\ref{eq21})}\\
\hline
 & 1 & 2 & 3 & 4 \\
\hline
    10.0&8.40-i1.40&8.50-i1.81 &8.48-i1.85 &8.49-i1.85 \\
\hline
    15.0&8.81-i1.80 &8.46-i1.87 &8.46-i1.85 &8.46-i1.85\\
\hline
    20.0&8.00-i2.18 &8.46-i1.85 &8.46-i1.85 &8.46-i1.85 \\
\hline
    30.0&8.64-i1.69 &8.47-i1.86 &8.47-i1.86 &8.47-i1.86 \\
\end{tabular}
\caption{\label{table5} Resonance positions of a 3-boson system 
 in the $J^{\pi}=0^+$ state interacting 
through  the S-wave interaction of Eq.(\ref{eq24}). The results are  without 
complex scaling but taking an increasing number of terms in 
the boundary condition (\ref{eq21}) into account. 
The correct energy eigenvalue is 
$E^{3B}_{res}=(8.47-i~1.86)$~MeV. Number of terms 1 means that only the  
$A(\theta)$ term has been used.}

\newpage

\begin{tabular}{|c|c|c|c|c|}   
\multicolumn{3}{c|}{$\lambda_V$} & $E_{2n}$ (MeV) & $E_{3n}$ (MeV)  \\ \hline
$^1S_0$    & $^3P_0$    & $^3P_2-^3F_2$ &       &                 \\ \hline
 1.8360    & 9.2804     & 5.2012        & -4.0   & -31.68       \\
 1.7241    & 8.9456     & 5.1422        & -3.0   & -28.06       \\
 1.5960    & 8.5851     & 5.0797        & -2.0   & -24.15        \\
 1.4370    & 8.1846     & 5.0123        & -1.0   & -19.73       \\
 1.3297    & 7.9584     & 4.9758        & -0.5   & -17.09    \\
 1.1930    & 7.7518     & 4.9442        & -0.1   & -14.21    \\
 1.1205    & 7.6981     & 4.9365        & -0.01  & -12.96    \\
 1.05      & 7.65       & 4.90          &  ~     & -11.13    \\ 
1.0        &  7.5       &  4.8 &  ~     &  -8.24   \\
1.0        &  1.0       &  4.8 &  ~     &  -7.12    \\ 
1.0        &  1.0       &  4.7 &  ~     &  -5.13    \\
1.0        &  1.0       &  4.6 &  ~     &  -3.29    \\
1.0        &  1.0       &  4.5 &  ~     &  -1.65    \\
1.0        &  1.0       &  4.4 &  ~     &  -0.23    \\
\end{tabular}
\caption{Required strength factors $\lambda_V$ for particular force components 
of the Gogny NN interaction together with the corresponding 
2- and 3-neutron ($3/2^-$) bound  energy eigenvalues.}\label{table6}

\pagebreak

\begin{tabular} {|c|c|c|c|c|c|c|}
$\lambda_{^3P_2-^3F_2}$&\multicolumn{6}{c|}{$E^{3n} (MeV)$ for particular number of terms in Eq.(\ref{eq21})}\\
\hline
 & 1 & 2 & 3 & 4 & 5 & 6\\
\hline
  4.30  &0.86 - i~0.068 &0.86 - i~0.068 &0.85 - i~0.092 &0.85 - i~0.092 &0.86 - i~0.068 &0.86 - i~0.068 \\
\hline
  4.25  &1.26 - i~0.27 &1.34 - i~0.21 &1.33 - i~0.21 &1.34 - i~0.21 &1.33 - i~0.22 &1.33 - i~0.22 \\
\hline
  4.20  &1.88 - i~0.38 &1.77 - i~0.41 &1.77 - i~0.41 &1.77 - i~0.41 &1.79 - i~0.38 &1.79 - i~0.38 \\
\hline
  4.15  &2.24 - i~0.53 &2.20 - i~0.62 &2.19 - i~0.62 &2.19 - i~0.63 &2.21 - i~0.64 &2.19 - i~0.63 \\
\hline
  4.10  &2.54 - i~0.70 &2.58 - i~0.87 &2.60 - i~0.86 &2.58 - i~0.88 &2.63 - i~0.89 &2.57 - i~0.84 \\
\hline
  4.05  &2.72 - i~0.90 &2.94 - i~1.15 &2.96 - i~1.16 &2.97 - i~1.14 &3.04 - i~1.15 &2.92 - i~1.18 \\
\hline
  4.00  &2.81 - i~1.08 &3.27 - i~1.49 &3.32 - i~1.46 &3.32 - i~1.46 &3.31 - i~1.47 &3.31 - i~1.47 \\
\hline
  3.95  &2.84 - i~1.20 &3.63 - i~1.86 &3.64 - i~1.81 &3.66 - i~1.79 &3.64 - i~1.80 &3.64 - i~1.80 \\
\hline
  3.90  &2.84 - i~1.30 &4.01 - i~2.24 &3.97 - i~2.16 &3.95 - i~2.16 &3.96 - i~2.14 &3.97 - i~2.14 \\
\hline
  3.85  &2.84 - i~1.37 &4.39 - i~2.55 &4.25 - i~2.53 &4.24 - i~2.53 &4.22 - i~2.54 &4.21 - i~2.51 \\
\hline
  3.80  &2.84 - i~1.41 &4.70 - i~2.82 &4.54 - i~2.91 &4.50 - i~2.92 &4.47 - i~2.94 &4.49  - i~2.93\\
\hline
  3.75  &2.84 - i~1.44 &4.88 - i~3.13 &4.77 - i~3.29 &4.76 - i~3.32 &4.73 - i~3.33 &4.78 - i~3.35 \\
\hline
  3.70  &2.84 - i~1.47 &5.07 - i~3.40 &4.99 - i~3.68 &4.95 - i~3.75 &4.98 - i~3.78 &4.96 - i~3.77 \\
\hline
  3.65  &2.84 - i~1.50 &5.17 - i~3.66 &5.12 - i~4.08 &5.14 - i~4.18 &5.16 - i~4.17 &5.14 - i~4.19 \\
\hline
  3.60  &2.84 - i~1.51 &5.23 - i~3.91 &5.21 - i~4.51 &5.30 - i~4.67 &5.31 - i~4.68 &5.33 - i~4.69 \\
\end{tabular}
\caption{\label{table7} The 3-neutron $3/2^-$ resonance state trajectories 
for the Gogny NN interaction obtained with increasing number of terms in 
the boundary condition (\ref{eq21}). 
The enhancement factors $\lambda$ for the 
$^1S_0$ and $^3P_0$ nn force components 
are equal one.}

\pagebreak

\begin{tabular} {|c|c|c|c|c|}
 $J^{\pi}$&\multicolumn{3}{c|}{$\lambda_V$} &  $E_{3n}$ (MeV) \\
\hline
 & $^1S_0$ & $^3P_0$ & $^3P_2-^3F_2$ &  \\
\hline
  $1/2^-$ &1.0 &5.5 &2.75 &2.82 - i~2.82 \\
\hline
  $3/2^-$ &1.0 &1.0 &3.70 &4.95 - i~3.75 \\
\hline
  $3/2^+$ &1.0 &5.0 &4.3 &5.74 - i~1.53 \\
\end{tabular}
\caption{\label{table8} The positions of 3-neutron resonances 
 for the Gogny potential obtained with four leading terms in 
the boundary condition (\ref{eq21}).}

\pagebreak

\begin{tabular}{|c|c|c|}  
 Re($\alpha$) & Im($\alpha$) & $E_{3n}$  \\
   ($^o$)  & ($^o$)     &  (MeV)     \\  \hline
10.0       &  0.0       &  2.88 - i~0.23 \\
20.0       &  0.0       &  2.88 - i~0.23 \\
30.0       &  0.0       &  3.11 - i~0.16 \\
\end{tabular}
\caption{CSM applied to the $J^{\pi}=3/2^+$ resonance state of the 
Gogny potential with enhancement factors $\lambda_i=(1.0, 5.3,4.6)$.   
 The correct energy eigenvalue 
is $E^{{3/2}^+}_{3n}=(2.88-i~0.23)$~MeV.}\label{table8a}

\pagebreak

\begin{tabular} {|c|c|c|c|c|}
 $J^{\pi}$&\multicolumn{3}{c|}{$\lambda_V$} &  $E_{3n}$ (MeV) \\
\hline
 & $^1S_0$ & $^3P_0$ & $^3P_2-^3F_2$ &  \\
\hline
  $3/2^-$ &1.0 &1.0 &3.25 &5.30 - i~3.53 \\
\hline
  $3/2^+$ &1.0 &4.6 &3.50 &5.93 - i~1.55 \\
\end{tabular}
\caption{\label{table9} The positions of 3-neutron resonances 
 for the Reid 93 potential obtained with four leading terms in 
the boundary condition (\ref{eq21}).}

\end{table}


\noindent
\begin{figure}
\caption{Analytical continuation of the $^3P_0$ 
Reid93 NN potential under the complex 
scaling transformation  $r \rightarrow  r~e^{i\alpha}$  
for five values of $\alpha= 0^o, 20^o, 40^o, 60^o$ and $80^o$. 
((a) real part, (b) imaginary part). \label{fig1}}
\end{figure}

\noindent
\begin{figure}
\caption{The real and imaginary parts of the reduced Faddeev amplitudes F 
for a 3-boson resonance in the state  $0^+$ without complex scaling 
((a) real part, (b) imaginary part).\label{fig2}}
\end{figure}

\noindent
\begin{figure}
\caption{The same as in Fig.\ref{fig2} but with  ECSM taking 
$\alpha=(40.0^o - i 40.0^o$) ((a) real part, (b) imaginary part).\label{fig3}}
\end{figure}

\noindent
\begin{figure}
\caption{The 3-neutron $1/2^-$, $3/2^-$, and $3/2^+$ 
resonance state trajectories 
for the Gogny NN interaction obtained with 4  terms in 
the asymptotic boundary condition. 
 Different points correspond to resonance positions 
with different values of the enhancement factors. For the $1/2^-$ state  
$\lambda_{^1S_0}=1.0$, $\lambda_{^3P_0}$ changes 
from 7.5 to 5.5 in steps of 0.1, and   
$\lambda_{^3P_2-3F_2}$ changes from 4.75 
to 2.75 in steps of 0.1. 
For the $3/2^-$ state  
$\lambda_{^1S_0}=1.0$, $\lambda_{^3P_0}=1.0$, and   
$\lambda_{^3P_2-3F_2}$ changes from 4.3 
to 3.7 in steps of 0.05.
 For the $3/2^+$ state  
$\lambda_{^1S_0}=1.0$, $\lambda_{^3P_0}$ changes 
from 5.5 to 5.0 in steps of 0.05, and   
$\lambda_{^3P_2-3F_2}$ changes from 4.8 
to 4.3 in steps of 0.05. 
\label{fig4}}
\end{figure}

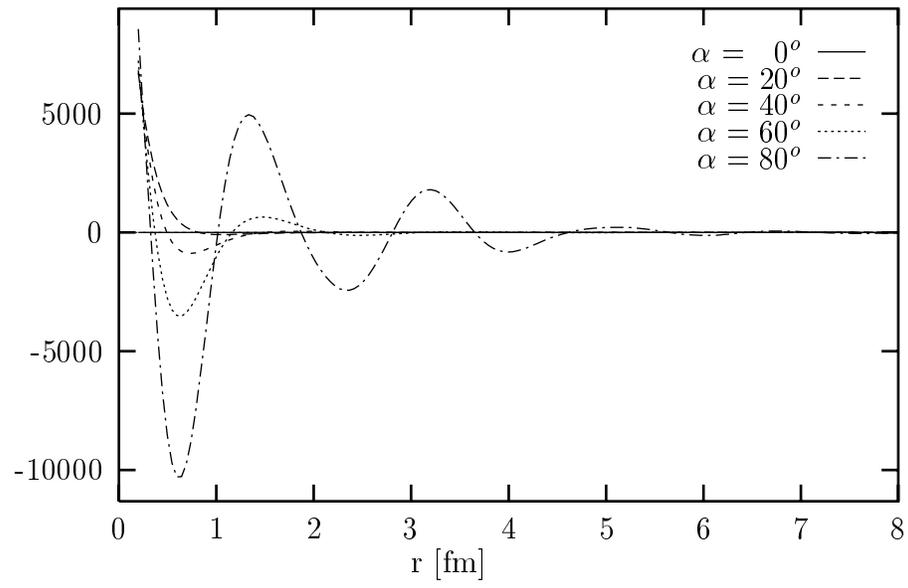
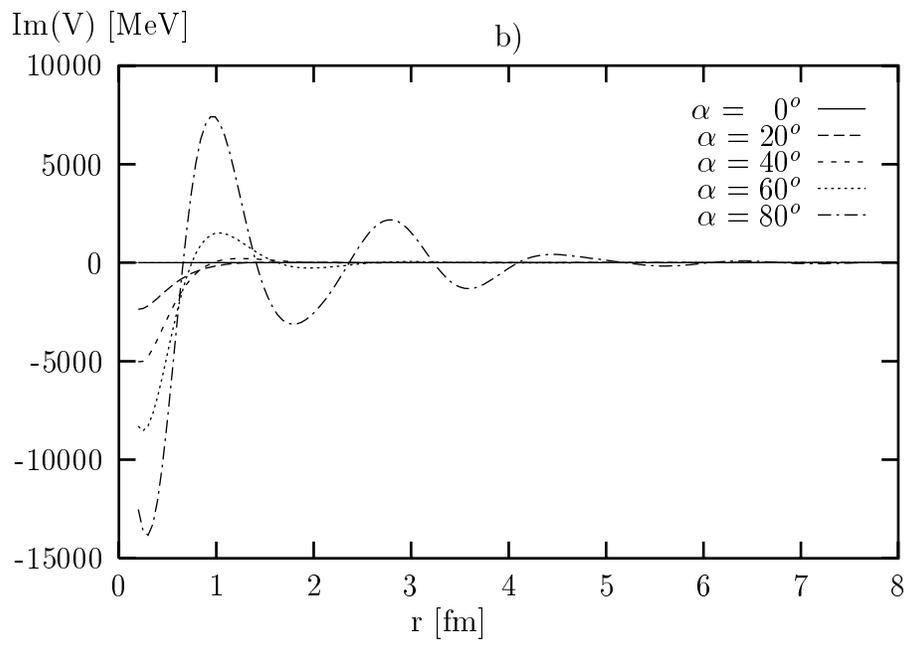
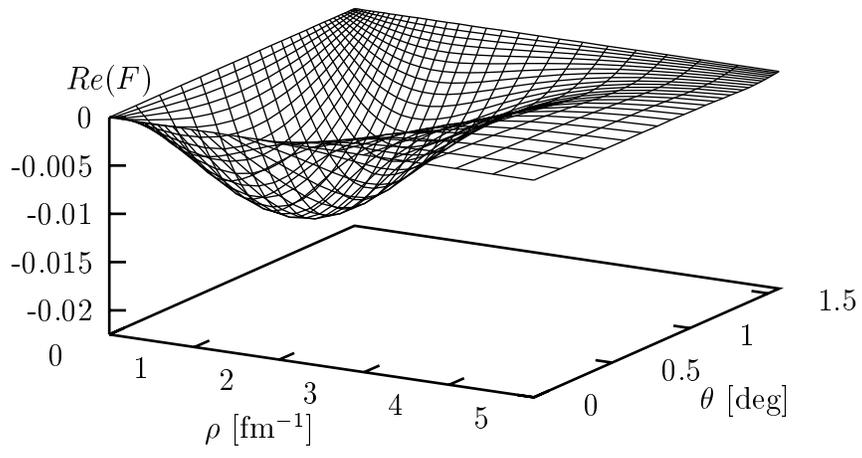
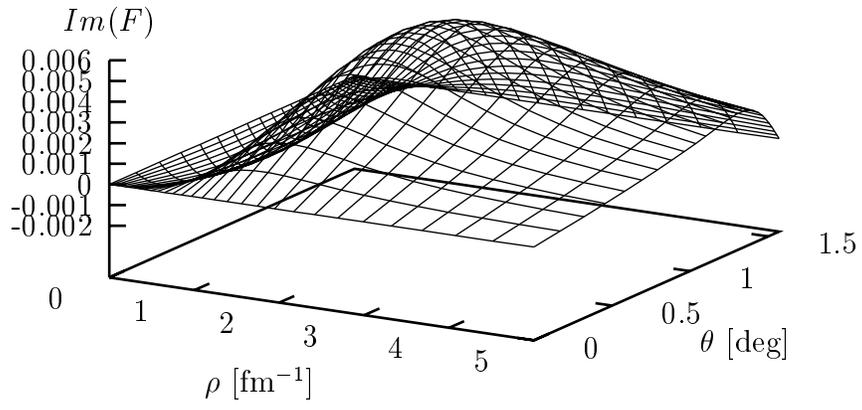
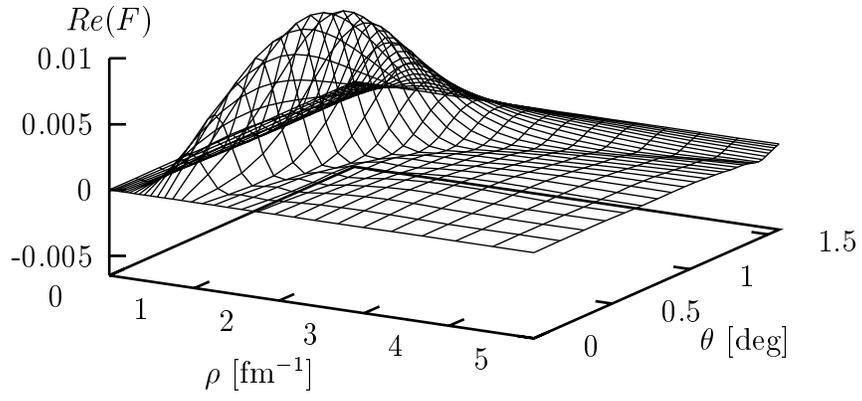
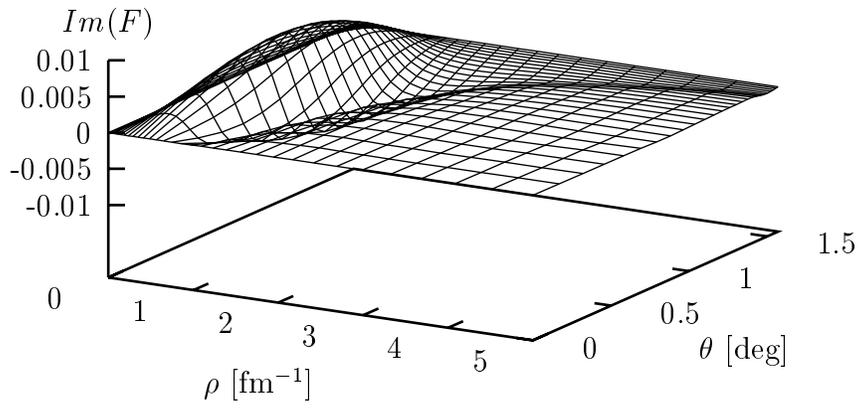
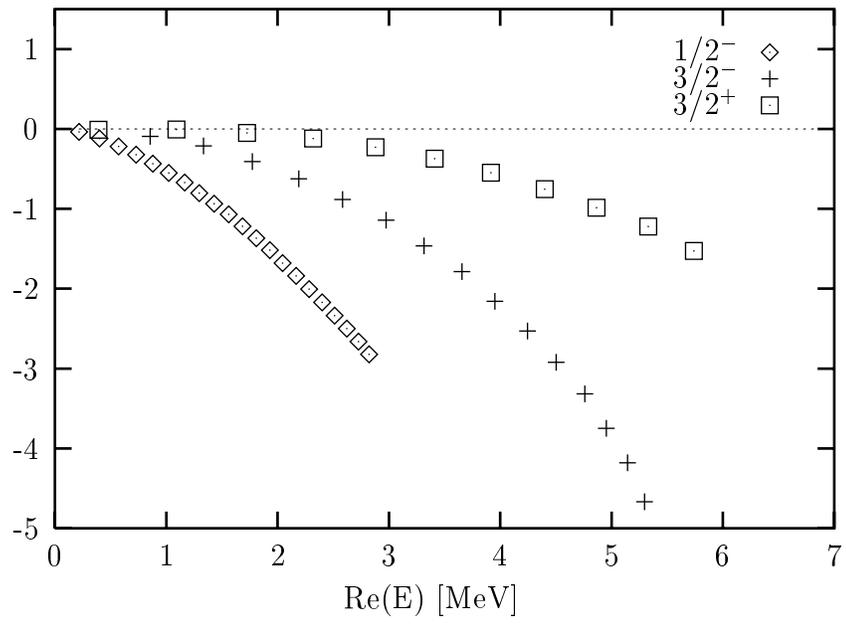
\end{document}